\documentclass[journal]{IEEEtran}
\usepackage{amsmath,amsfonts}
\usepackage{algorithmic}
\usepackage{array}
\usepackage[caption=false,font=normalsize,labelfont=sf,textfont=sf]{subfig}
\usepackage{textcomp}
\usepackage{stfloats}
\usepackage{url}
\usepackage{verbatim}
\usepackage{graphicx}
\usepackage{mathrsfs}
\usepackage{balance}
\usepackage[noadjust]{cite}
\usepackage{booktabs}

\usepackage[]{hyperref}

\usepackage{xcolor}
\usepackage{tikz}

\providecommand{\I}{\ensuremath{\mathrm{i}}}        
\providecommand{\RE}{\ensuremath{\mathrm{Re}}}  

\providecommand{\D}{\,\mathrm{d}}               
\providecommand{\M}[1]{\mathbf{#1}}             
\providecommand{\T}[1]{\mathrm{#1}}             

\providecommand{\TARC}{\Gamma^\T{t}}     

\providecommand{\herm}{\mathrm{H}}

\providecommand{\Ivec}{\ensuremath{\M{I}}}
\providecommand{\IvecH}{\Ivec^{\herm}}
\providecommand{\Vvec}{\ensuremath{\M{V}}}

\providecommand{\pVvec}{\ensuremath{\M{v}}}
\providecommand{\pVvecH}{\ensuremath{\pVvec^\herm}}

	\providecommand{\PradDD}{\dfrac{\D^2\Prad}{\D\omega^2}}
	
	\providecommand{\PinDD}{\dfrac{\D^2\Pin}{\D\omega^2}}
	
	\providecommand{\etaDD}{\dfrac{\D^2\eta}{\D\omega^2}}
	
	\providecommand{\YmatFS}{\ensuremath{\M{Y}}_0}
	
	\providecommand{\RmatFSMoM}{{\ensuremath{\M{R}}}_0}
	\providecommand{\XmatFSMoM}{{\ensuremath{\M{X}}}_0}
	\providecommand{\ZmatFSMoM}{{\ensuremath{\M{Z}}}_0}

	\providecommand{\YmatFSMoM}{{\ensuremath{\M{Y}}}_0}
	\providecommand{\WmatMoM}{{\ensuremath{\M{W}}}}
	
	\providecommand{\RmatL}{\ensuremath{\M{R}}_{\rho}}

	\providecommand{\Lambmat}{\ensuremath{\M{\Lambda}}}
	\providecommand{\Dmat}{\ensuremath{\M{D}}}
	\providecommand{\Pmat}{\ensuremath{\M{P}}}

	\providecommand{\Kmat}{\ensuremath{\M{k}}}

	\providecommand{\GmatDD}{\ensuremath{\dfrac{\partial^2\pGmat_0}{\partial\omega^2}}}

	\providecommand{\YmatD}{\ensuremath{\dfrac{\partial\pYmat}{\partial\omega}}}
	\providecommand{\YmatDD}{\ensuremath{\dfrac{\partial^2\pYmat}{\partial\omega^2}}}
	\providecommand{\YmatHD}{\ensuremath{\dfrac{\partial\pYmat^\herm}{\partial\omega}}}
	\providecommand{\YmatHDD}{\ensuremath{\dfrac{\partial^2\pYmat^\herm}{\partial\omega^2}}}

	\providecommand{\pGmat}{\ensuremath{\M{g}}}
	\providecommand{\pBmat}{\ensuremath{\M{b}}}
	\providecommand{\pYmat}{\ensuremath{\M{y}}}
	\providecommand{\pWmat}{\ensuremath{\M{w}}}
	
	\providecommand{\pGmatFS}{\ensuremath{\M{g}}_0}
	\providecommand{\pBmatFS}{\ensuremath{\M{b}}_0}
	\providecommand{\pYmatFS}{\ensuremath{\M{y}}_0}

	\providecommand{\pKimat}{\ensuremath{\Kmat}_\T{i}}
	\providecommand{\pKrmat}{\ensuremath{\M{k}_\T{r}}}
	
	\providecommand{\pAvec}{\ensuremath{\M{a}}}
	
	\providecommand{\pBvec}{\ensuremath{\M{b}}}

	\providecommand{\Prad}{P_\T{rad}}
	\providecommand{\Pin}{P_\T{in}}
	\providecommand{\Preact}{P_\T{react}}

	\providecommand{\We}{\ensuremath{W_\mathrm{e}}}
	\providecommand{\Wm}{\ensuremath{W_\mathrm{m}}}
	
	\newcommand{\ie}{\textit{i.e.}{}}

	\newcommand{\BE}{\begin{equation}}
		\newcommand{\EE}{\end{equation}}
	\newcommand{\BEn}{\begin{equation*}}
		\newcommand{\EEn}{\end{equation*}}
	\newcommand{\BF}{\begin{figure}\centering}
		\newcommand{\EF}{\end{figure}}
	\newcommand{\BT}{\begin{table}\centering}
		\newcommand{\ET}{\end{table}}

	\RequirePackage[nolist]{acronym}
	\newacro{MoM}[MoM]{method of moments}
	\newacro{FEM}[FEM]{finite element method}
	\newacro{FDTD}[FDTD]{finite-difference time-domain method}
	\newacro{MOO}[MOO]{multiobjective optimization}
	\newacro{CM}[CM]{characteristic mode}
	\newacro{PEC}[PEC]{perfect electric conductor}
	\newacro{PMC}[PMC]{perfect magnetic conductor}
	\newacro{EP}[EP]{eigenvalue problem}
	\newacro{GEP}[GEP]{generalized eigenvalue problem}
	\newacro{EFIE}[EFIE]{electric field integral equation}
	\newacro{SVD}[SVD]{singular value decomposition}
	\newacro{RWG}[RWG]{Rao-Wilton-Glisson}
	\newacro{EM}[EM]{electromagnetic}
	\newacro{DOF}[DOF]{\mbox{degrees-of-freedom}}
	\newacro{VNA}[VNA]{\mbox{vector network analyzer}}
	\newacro{TARC}[TARC]{total active reflection coefficient}
	
\begin{document}
\title{Multiport Antenna Q-factor}
\author{Vojtech Neuman, Miloslav Capek, \IEEEmembership{Senior Member, IEEE}, Lukas Jelinek
\thanks{Manuscript received \today; revised \today. The Czech Science Foundation supported this work under project~\mbox{No.~24-11678S}.}
\thanks{V. Neuman, M. Capek and L. Jelinek are with the Czech Technical University in Prague, Prague, Czech Republic (e-mails: \{vojtech.neuman; miloslav.capek; lukas.jelinek\}@fel.cvut.cz).}
\thanks{Color versions of one or more of the figures in this paper are available online at http://ieeexplore.ieee.org}
}


\maketitle

\begin{abstract}
This article proposes an estimate of multiport antenna bandwidth based on a generalization of a single-port Q-factor. The explicit derivation is based on converting the stored energy matrix to its port equivalent and on the port parameters themselves. The work discusses the bandwidth dependencies on feeding and matching. Derived formulas are shown to utilize the total active reflection coefficient and allow for a single-frequency bandwidth evaluation. Examples comprising two different dipole arrays and electrically large patch antenna arrays validate the theory. 
\end{abstract}


\section{Introduction}
\label{sec:Intro}
\IEEEPARstart{A}{ntenna} bandwidth represents an important figure of merit~\cite{book:BalanisAntTheory} directly connected to the channel capacity of a wireless link~\cite{book:MolischWireComm}. In the case of narrowband antennas~\cite{book:BalanisAntTheory, book:VolakisSmallAnt}, bandwidth is commonly estimated using its inverse proportionality to the Q-factor~\cite{art:Yaghjian2025FundAntBandwidth}, which can be considered as an equivalent figure of merit having the advantage of a single frequency evaluation.

Research on Q-factor has a rich history~\cite{book:VolakisSmallAnt, art:Yaghjian2025FundAntBandwidth} and remains a popular topic in the scientific spotlight~\cite{art:Yaghjian2025FundAntBandwidth, art:Passalacqua2023QBoundMaxDirectivity, art:Lundgren2025FundLimitCharModesSlopes}. Its origins lie in mechanical oscillations~\cite{book:FeynmanLecturesVolI} and the Q-factor was later introduced into electrical engineering~\cite{art:Smith1986OriginsOfQualityFactor} with pioneering work on antennas being carried out by Wheeler~\cite{art:Wheeler1947FundLimitSmallAnt} who first observed the relation between antenna volume and its bandwidth. The small antenna theory was further developed by follow-up papers focusing on spherical mode expansion into equivalent circuits~\cite{art:Chu1948PhysLimitOmniDirAntenna, art:Harrington1960EffectAntSize,art:Thal2006RadiationQSphericalWireAntennas}, field-based techniques~\cite{art:Collin1964EvalAntQ, art:Fante1969QualityFactorGenIdealAntenna, art:McLean1996ReexaminationFundLimit, art:Geyi2003MethodEvalQ} or far-fields~\cite{art:Thiele2003OnLowerBoundRadQ}, studying the relation of the Q-factor, or so called radiation Q-factor, with stored energy and established lower limits. A well-known result is the lowest Q-factor that occurs in a combination of TE and TM modes~\cite{art:Harrington1960EffectAntSize, art:Fante1969QualityFactorGenIdealAntenna}, which led Best to publish works about spherical helix antennas~\cite{art:Best2004RadPropertiosHelixAntennas}, a result later refined in~\cite{art:Kim2012MinimumQ} to analytically show the optimal shape being a loxodrome~\cite{art:Kim2012MinimumQ,art:Tucek2023DensityBasedTopoOpt}. 

Stored energy is a topic related to the Q-factor~\cite{art:Schab2018EnergyStore}, where Harrington~\cite{art:Harrington1972ControlRadScat} presented a way of calculating stored energy consisting of the frequency derivative of the imaginary part of the impedance matrix obtained by \ac{MoM}~\cite{book:HarringtonFieldComp}. This approach was proven valid in~\cite{art:Vandenbosch2010ReactEnergyImp} based on relations developed by Yaghjian and Best~\cite{art:Yaghjian2003ImpBandAnt}, as long as the stored energy matrix is positive definite~\cite{art:Gustafsson2016AntCurrOpt}. Numerical evaluations have opened a new avenue for Q-factor computations~\cite{art:Gustafsson2016AntCurrOpt}, leading to a methodology for establishing fundamental bounds for arbitrary shapes~\cite{art:Capek2017MinAntQ}, which confirmed the previous results of optimal mode composition~\cite{art:Capek2016OptCompModalCurrQ}. These calculations were extended to the substructure fundamental bounds~\cite{art:Schab2018LowerBoundSubstructQ, art:Jonsson2021OptBandwidthPosition}, inclusion of dielectric materials~\cite{art:Nel2023QFactorBoundMPA}, and a study of the role of the symmetries~\cite{art:Capek2020RoleSymFundBounds} on results. The Q-factor numerical evaluation enabled automated design schemes based on topology optimization~\cite{art:Tucek2023DensityBasedTopoOpt, art:Ethier2014AntennaShapeSynthWithoutFeedpoint, art:Cismasu2014AntBandwithOpt}.

Work~\cite{art:Fante1969QualityFactorGenIdealAntenna} offered a first direct relationship between the conductance fractional bandwidth and the Q-factor value, ultimately leading to a new branch of research based on the examination of antenna impedance (so-called impedance Q-factor). By utilizing Foster's reactance theorem, Geyi~\cite{art:Geyi2000FosterReactanceRadiationQ} proposed a way to calculate the Q-factor by the differentiation of the reactance. Work~\cite{art:Best2004FosterReactanceTheorem} alluded to problematic spots with the approach for antennas operating near anti-resonance, which were later carried out in~\cite{art:Yaghjian2003ImpBandAnt} with a concept of the impedance bandwidth and impedance Q-factor, and, importantly, a new formula for the accurate prediction of fractional bandwidth from the Q-factor with a given reflection coefficient limit. However, the impedance Q-factor may attain zero values~\cite{art:Gustafsson2006BandwidthQResonance} and, as with other Q-factor definitions, is limited in the bandwidth prediction of multiple close resonances~\cite{art:Stuart2007LimitRelatQfactDoubRes, art:Capek2015OnFuncRelBetweenQfact}. Later research focused on alternative bandwidth estimations~\cite{art:Vorobyev2010ConductanceBandwidth} and the inclusion of far-field properties in the calculation~\cite{art:Thal2015AntennaQCalculationRadModes}. The relation between the radiation Q-factor~\cite{art:Collin1964EvalAntQ} and impedance Q-factor~\cite{art:Yaghjian2003ImpBandAnt} was shown in~\cite{art:Capek2014MeasQfactObsEnergies, misc:Gustafsson2014QfactAntDispersive}, and the bandwidth predictions were extensively compared in~\cite{art:Gustafsson2015AntennaQStoredEnergy}. The impedance Q-factor was revisited in~\cite{art:Yaghjian2025FundAntBandwidth}, which contains a treatment of multiple close resonances.

So far, less attention has been given to the Q-factor of multiport antennas. The bandwidth itself is more complex to establish~\cite{art:Lau2006ImpactMatchNetwork, art:Nie2017BandwidthAnalysisMultiportI} and depends heavily on antenna feeding. The definition used throughout this article comes from~\cite{art:Manteghi2005MultiCharWideBand}. Nevertheless, the introduced theory can also use transducer efficiencies~\cite {art:Broyde2022RadTransEfficiencies}. One of the first works dealing with the Q-factor of multiple-port antennas is~\cite{art:Foltz1998BandwidthLimitMultiIsolatedPort}, which did not consider mutual coupling across ports~\cite{art:Stein1962OnCrossCoupling}. This deficiency was addressed in~\cite{art:Sten2008ImpedanceQualityFactorMultiport} where the Q-factor was derived with dependence on the multiport antenna excitation. Work~\cite{art:Wang2010BandwidthEnhancmentMutualCoupling} took the effect of the feeding network into account but did not provide the formula with feeding coefficients. A method based on the modal analysis of decoupled ports was introduced in~\cite{art:Yang2016SystShapeOptCharModes}. Following the idea of the impedance Q-factor~\cite{art:Yaghjian2003ImpBandAnt}, paper~\cite{art:Luomaniemi2021QfactMultiAntenna} presented a multiport Q-factor definition considering port excitation and port matrices.

The presented article builds upon the results of previous work~\cite{art:Yaghjian2003ImpBandAnt, art:Capek2020FindOptTotalRefCoef} and develops a generalized impedance Q-factor for multiport antennas with explicit dependence on matching and feeding coefficients as bilinear forms. The radiation Q-factor is expressed in terms of stored energies, and the generalized impedance Q-factor is derived from the total active reflection coefficient~\cite{art:Manteghi2005MultiCharWideBand}. It is shown that the generalized Q-factor can be reduced back to the impedance Q-factor~\cite{art:Yaghjian2003ImpBandAnt} and that results from~\cite{art:Luomaniemi2021QfactMultiAntenna} represent a special case. The article proposes a formula for estimating fractional bandwidth and discusses properties with respect to single-frequency evaluations and the incorporation of losses. A series of examples compares the proposed Q-factors with~\cite{art:Luomaniemi2021QfactMultiAntenna}, and shows that the generalized impedance Q-factor also works for electrically larger antennas. 

The rest of this article is organized as follows. Section~\ref{sec:Bandwidth} establishes the multiport antenna bandwidth and Q-factor terminology. Next, in Section~\ref{sec:PortOperators}, the reduction of \ac{MoM} operators to their port matrices is shown and applied to the radiation Q-factor. Section~\ref{sec:TARCApproximation} develops bandwidth approximation by expanding total efficiency into a Taylor polynomial. Sections~\ref{sec:ExampleDipole2Port}, \ref{sec:ExampleDipole4Port}, and~\ref{sec:ExamplePatchArray} examine the developed formulas on a series of examples. The article concludes in Section~\ref{sec:Conclusion}.

\section{Multiport Antenna Bandwidth}
\label{sec:Bandwidth}
Throughout the article, we assume a time-harmonic steady state at angular frequency $\omega$ with time derivatives replaced with $\partial/\partial t \rightarrow \I\omega$ where $\I$ is the imaginary unit. For multiport antennas, a well-defined measure of the performance is the \ac{TARC} which is defined as~\cite{art:Manteghi2005MultiCharWideBand}
\begin{equation}
\TARC = \sqrt{1 - \eta}, \label{eq:TARC}
\end{equation}
where $\eta$ represents total antenna efficiency. In the case of lossless single-port antennas, the \ac{TARC} is simplified to the magnitude of the reflection coefficient~\cite{book:PozarMicroEng}. Bandwidth~$B$ of the antenna~\cite{art:IEEE1969StdDefAnt} can then be established as a range of frequencies 
\begin{equation}
B = \left\{\omega \in \mathbb{R}\,\middle\vert\, \TARC\left(\omega\right) < \TARC_\T{max}\right\},
\end{equation}
where $\TARC_\T{max}$ represents the maximal allowed \ac{TARC} level. 

Considering electrically small antennas with a single resonance at $\omega_0$, absolute bandwidth $B$ is often replaced with the fractional bandwidth
\begin{equation}
F = \frac{\omega_\T{+} - \omega_\T{-}}{\omega_0}, \label{eq:FBW}
\end{equation}
where $\omega_\T{+}$ and $\omega_\T{-}$ represent the upper and lower limits of bandwidth. Fractional bandwidth is inversely proportional to the Q-factor~\cite{book:VolakisSmallAnt}
\begin{equation}
F \propto \frac{1}{Q},
\end{equation}
which can be used as a single-frequency bandwidth estimate. There are two classes of methods for determining the Q-factor. The first one operates with stored energies and is denoted as the radiation Q-factor~\cite{art:Vandenbosch2010ReactEnergyImp}
\begin{equation}
Q_\T{rad} = \frac{2\omega_0\max{\left\{\Wm,\We\right\}}}{\Prad}, \label{eq:RadiationQ}
\end{equation}
where $\Wm$ and $\We$ are the cycle-mean stored magnetic and electric energies, respectively, and $\Prad$ represents the cycle-mean radiated power. The second class is derived from the antenna input impedance~\cite{art:Yaghjian2003ImpBandAnt}
\begin{equation}
Q_\T{Z} = \frac{\omega_0\left\vert Z'\left(\omega_0\right)\right\vert}{2R\left(\omega_0\right)}, \label{eq:ImpedanceQ}
\end{equation}
where $Z = R + \I X$ is the input impedance of the antenna, and is referred to as the impedance Q-factor. Both relations have a strong physical connection~\cite{art:Yaghjian2003ImpBandAnt} and, therefore, return similar predictions~\cite{misc:Gustafsson2014QfactAntDispersive}, as long as the initial assumptions are not violated.

The rest of this article is aimed at developing relations for the radiation Q-factor~\eqref{eq:RadiationQ} and the multiport generalization of the impedance Q-factor~\eqref{eq:ImpedanceQ}. Both introduced formulas are compared on a series of examples.

\section{Antenna Characterization on Port Level}
\label{sec:PortOperators}
Considering integral equations~\cite{book:ChewIntEqua,book:VolakisIntEqua} for electromagnetic modeling, a lossless antenna\footnote{The lossless antennas are assumed only for the simplicity of the derivations, and the effect of the losses will be discussed in the following section.} can be described with an impedance matrix
\begin{equation}
\ZmatFSMoM\Ivec = \Vvec, \label{eq:ImpedanceMatrix}
\end{equation}
where $\Ivec \in \mathbb{C}^{N\times 1}$ represents a vector of current expansion coefficients, $N$ denotes the number of basis functions, see article~\cite{book:HarringtonFieldComp} for details, and $\Vvec$ is a vector describing the feeding of the antenna. The impedance matrix consists of real and imaginary parts
\begin{equation}
\ZmatFSMoM = \RmatFSMoM + \I\XmatFSMoM,
\end{equation}
and, together with the current expansion vector, can be used to evaluate cycle-mean radiated power~\cite{book:HarringtonFieldComp}
\begin{equation}
\Prad = \frac{1}{2}\IvecH\RmatFSMoM\Ivec,
\end{equation}
and cycle-mean reactive power
\begin{equation}
\Preact = \frac{1}{2}\IvecH\XmatFSMoM\Ivec,
\end{equation}
respectively. Following~\cite{art:Vandenbosch2010ReactEnergyImp}, the stored energy matrix reads 
\begin{equation}
\WmatMoM = \frac{\partial \XmatFSMoM}{\partial \omega},
\end{equation}
which is used to evaluate magnetic and electric energies
\begin{align}
\Wm = \frac{1}{8\omega}\IvecH\left(\omega\WmatMoM + \XmatFSMoM\right)\Ivec, \\
\We = \frac{1}{8\omega}\IvecH\left(\omega\WmatMoM - \XmatFSMoM\right)\Ivec,
\end{align}
respectively~\cite{art:Capek2017MinAntQ}. Taking into account the properties of $\max$ function~\cite{art:Gustafsson2016AntCurrOpt}, the radiation Q-factor evaluation can be carried out as
\begin{equation}
Q_\T{rad} = \frac{\omega}{2}\frac{\IvecH\WmatMoM\Ivec}{\IvecH\RmatFSMoM\Ivec} + \frac{1}{2}\frac{\left\vert\IvecH\XmatFSMoM\Ivec\right\vert}{\IvecH\RmatFSMoM\Ivec}. \label{eq:RadiationQMoM}
\end{equation}

Let current vector $\Ivec$ be excited by $P$ discrete voltage ports~\cite{book:Gibson2021MoM}. The method introduced in~\cite{art:Capek2020FindOptTotalRefCoef} is adopted here to convert \ac{MoM} matrices to their port equivalents. We start with the definition of the voltage vector
\begin{equation}
\Vvec = \Dmat\Pmat\pVvec,
\end{equation}
using a port vector of voltage excitation coefficients~$\pVvec$. The indexing binary matrix~$\Pmat \in \mathbb{B}^{N \times P}$ is given by
\begin{equation}
P_{np} = 
\begin{cases}
1 & \text{$p$-th port is placed at $n$-th position}, \\
0 & \text{otheriwse},
\end{cases}
\end{equation}
and matrix $\Dmat$ is a diagonal matrix ensuring\footnote{For commonly used \ac{RWG} basis functions~\cite{art:Rao1982EleScattSurf}, matrix $\Dmat$ is a diagonal matrix with the edge lengths associated with common \ac{RWG} edges.} proper physical units~\cite{art:Capek2020FindOptTotalRefCoef}. In the article, we assume port voltages as frequency-independent. The excited current is
\begin{equation}
\Ivec = \left(\ZmatFSMoM\right)^{-1}\Vvec = \YmatFSMoM\Dmat\Pmat\pVvec,
\end{equation}
with $\YmatFSMoM$ being the admittance matrix. A bilinear form in the current vector $\Ivec$ can then be transformed into its port equivalent as
\begin{equation}
\IvecH{\M{M}}\Ivec = \pVvecH\Pmat^\herm\Dmat^\herm\YmatFSMoM^\herm{\M{M}}\,\YmatFSMoM\Dmat\Pmat\pVvec = \pVvecH\M{m}\pVvec,
\end{equation}
where $\M{m}$ represents the former matrix\footnote{All matrices denoted with uppercase letters belong to the level of the basis functions discretizing the integral equation~\eqref{eq:ImpedanceMatrix}, while the lowercase is used for port matrices.} in reduced basis~\cite{art:Capek2020FindOptTotalRefCoef}. Continuing with this strategy allows us to rewrite formula~\eqref{eq:RadiationQMoM} with port voltages
\begin{equation}
Q_\T{rad} = \frac{\omega}{2}\frac{\pVvecH\pWmat\pVvec}{\pVvecH\pGmatFS\pVvec} + \frac{1}{2}\frac{\left\vert\pVvecH\pBmatFS\pVvec\right\vert}{\pVvecH\pGmatFS\pVvec}, \label{eq:RadiationQPort}
\end{equation}
where $\pGmatFS$ and $\pBmatFS$ are real and imaginary parts, respectively, of the port admittance matrix
\begin{equation}
\pYmatFS = \pGmatFS + \I\pBmatFS.
\end{equation}
This gives a radiation Q-factor of a multiport antenna. Obtaining the equivalent of the impedance Q-factor for a multiport antenna is more involved and is shown in the next section.

\section{First-Order TARC Approximation}
\label{sec:TARCApproximation}
\begin{figure}[t]
\centering
\includegraphics[scale=1]{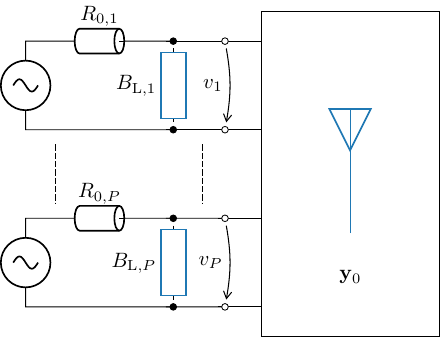}
\caption{Antenna system with $P$ ports with voltages $v_p$, each being connected to a transmission line with line impedance $R_{0,p}$ and parallel tuning element $B_{\T{L},p}$, together representing the input impedances. Matrix $\pYmatFS$ represents the antenna admittance matrix.}\label{pic:MultiPortAntennaImp}
\end{figure}

We start with a multiport antenna whose ports are connected to ideal transmission lines of real-valued impedances~$R_{0,n}$ and parallel tunning susceptances~$B_{\T{L},n}$, and the values of both are used to match the antenna, see Fig.~\ref{pic:MultiPortAntennaImp}.

Specifically, the incident power waves~\cite{book:PozarMicroEng} collected in vector $\pAvec$ and the reflected power waves collected in vector $\pBvec$ are expressed in terms of port voltages as~\cite{art:Capek2020FindOptTotalRefCoef}
\begin{align}
\pAvec = \pKimat\pVvec = \frac{1}{2}\left(\Lambmat^{-1} + \Lambmat\left(\pYmat_0 + \I\pBmat_\T{L}\right)\right)\pVvec, \label{eq:Avec} \\
\pBvec = \pKrmat\pVvec = \frac{1}{2}\left(\Lambmat^{-1} - \Lambmat\left(\pYmat_0 + \I\pBmat_\T{L}\right)\right)\pVvec, \label{eq:Bvec}
\end{align}
with $\Lambmat$ being the diagonal matrix expressing generator impedances as
\begin{equation}
\Lambda_{nn} = \sqrt{R_{0,n}},
\end{equation}
and $\pBmat_\T{L}$ being the diagonal matrix of the tuning susceptances. 

Analogously to the derivation of the impedance Q-factor of a single-port antenna~\cite{art:Yaghjian2003ImpBandAnt, art:Yaghjian2025FundAntBandwidth}, we begin with the assumption of a lossless and matched antenna with vanishing reflected power waves~$\pBvec = \M{0}$ and total efficiency equal to unity at a specific frequency~$\omega_0$. We also note that such conditions require specific incident waves~$\pAvec$ as dictated by relation~\eqref{eq:Avec}.

The next step is the expansion of the antenna efficiency $\eta\left(\omega\right)$ around its maximum at frequency~$\ \omega_0$. The expansion reads 
\begin{equation}
\eta\left(\omega\right) = 1 + \frac{1}{2}\etaDD\bigg|_{\omega_0}\left(\omega - \omega_0\right)^2 + O(\omega^3), \label{eq:EfficiencyTaylorExpansion}
\end{equation}
where $O\left(\omega^3\right)$ describes the cubic error of the approximation. Substituting in the relation~\eqref{eq:TARC} results in
\begin{equation}
\TARC\left(\omega\right) \approx \frac{\left\vert\omega - \omega_0 \right\vert}{\omega_0}\cdot\sqrt{-\frac{\omega_0^2}{2}\etaDD\bigg|_{\omega_0}}, \label{eq:TARCApprox}
\end{equation}
where the square root term is called the \ac{TARC} Q-factor
\begin{equation}
Q_\T{\TARC} = \sqrt{-\frac{\omega_0^2}{2}\etaDD\bigg|_{\omega_0}}. \label{eq:TARCQ}
\end{equation}
Later sections will show that this \ac{TARC} Q-factor is a good estimate of the relative bandwidth of a multiport antenna.

In order to evaluate \ac{TARC} Q-factor efficiently, total efficiency $\eta$ is written as
\begin{equation}
\eta\left(\omega\right) = \frac{\Prad\left(\omega\right)}{\Pin\left(\omega\right)},
\end{equation}
which, taking into account the lossless antenna matched at $\omega_0$, allows us to evaluate its second derivative as
\begin{equation}
\etaDD\bigg|_{\omega_0} = \dfrac{1}{\Pin}\PradDD\bigg|_{\omega_0} - \dfrac{1}{\Pin}\PinDD\bigg|_{\omega_0}. \label{eq:EtaSecDerivative}
\end{equation}
Upon substitution from~\eqref{eq:Avec}, the frequency derivative of the incident power~$\Pin = \pAvec^\herm \pAvec / 2$ gives
\begin{multline}
\PinDD = \frac{1}{8}\pVvecH\left(\YmatDD + \YmatHDD\right)\pVvec  \\ + \frac{1}{8}\pVvecH\left(2\YmatHD\Lambmat^2\YmatD\right)\pVvec \\ + \frac{1}{8}\pVvecH\left(\YmatHDD\Lambmat^2\pYmat  + \pYmat^\herm\Lambmat^2\YmatDD\right)\pVvec,
\end{multline}
with~$\pYmat = \pYmat_0 + \I\pBmat_\T{L}$.
Proceeding with the second derivative of radiated power, we obtain
\begin{equation}
\PradDD = \frac{1}{2}\pVvecH\GmatDD\pVvec = \frac{1}{4}\pVvecH\left(\YmatDD + \YmatHDD\right)\pVvec,
\end{equation}
and substituting all back to~\eqref{eq:EtaSecDerivative}, the final relation reads

\begin{multline}
\etaDD\bigg|_{\omega_0} = - \frac{1}{2}\dfrac{\pVvecH\YmatHD\Lambmat^2\YmatD\pVvec}{\pVvecH\pKimat^\herm\pKimat\pVvec}  \\[3pt]   +\frac{1}{2}\dfrac{\RE\left\{\pVvecH\YmatHDD \Lambmat \left(\Lambmat^{-1} -\Lambmat\pYmat\right)\pVvec\right\}}{\pVvecH\pKimat^\herm\pKimat\pVvec}. \label{eq:TotEffDDFull}
\end{multline}
If the antenna is matched\footnote{Note that the assumption of a reflectionless antenna forbids the constant incident power at all frequencies. At a single frequency, this normalization is, however, allowed.} ($\pBvec = \M{0}$), the second term in~\eqref{eq:TotEffDDFull} is equal to zero as is seen from relation~\eqref{eq:Bvec}. Finally, the \ac{TARC} Q-factor is given by
\begin{equation}
Q_\T{\TARC} = \sqrt{\frac{\omega_0^2}{4}\frac{\pVvecH\YmatHD\Lambmat^2\YmatD\pVvec}{\pVvecH\pKimat^\herm\pKimat\pVvec}} = \frac{\omega_0}{2}\frac{\left\Vert\Lambmat\YmatD\pVvec\right\Vert}{\left\Vert\pKimat\pVvec\right\Vert}\label{eq:QfactorTARCAdmittance}
\end{equation}
with derivatives evaluated at $\omega_0$, and the maximal allowable \ac{TARC} defined as
\begin{equation}
\TARC_\T{max} = \frac{\left\vert \omega_\pm - \omega_0\right\vert}{\omega_0}Q_\TARC,
\end{equation}
with the fractional bandwidth given as
\begin{equation}
F \approx \frac{2\TARC_\T{max}}{Q_{\TARC}} \label{eq:FBWApproxQ}.
\end{equation}

As an important side note and a check of consistency, it is noted that~\eqref{eq:QfactorTARCAdmittance} reduces to~\eqref{eq:ImpedanceQ} for a lossless matched single-port antenna. To see this, notice that for a single port and matched antenna, there is
\begin{equation}
\dfrac{\partial Y}{\partial \omega} = - \dfrac{1}{Z^2} \dfrac{\partial Z}{\partial \omega} = - \dfrac{1}{R_0^2} \dfrac{\partial Z}{\partial \omega}.
\end{equation}
where line impedance $R_0 = Z\left(\omega_0\right)$. Furthermore, the connected voltage reduces to a scalar value, which cancels itself. Substituting into~\eqref{eq:QfactorTARCAdmittance} gives
\begin{equation}
Q_\T{\TARC} = \frac{\omega_0}{2}\left\vert\dfrac{\sqrt{R_0}\dfrac{\partial Y}{\partial \omega}}{\dfrac{1 + R_0Y}{2\sqrt{R_0}}}\right\vert = \frac{\omega_0}{2R_0}\left\vert \dfrac{\partial Z}{\partial \omega} \right\vert = Q_\T{Z}.
\end{equation}

As with impedance Q-factor~$Q_\T{Z}$, the evaluation of \ac{TARC} Q-factor~$Q_\T{\TARC}$ can be based on measurement\footnote{The analysis requires only port parameters, with the Q-factor subsequently evaluated through post-processing.} or advantageously on the matrices readily available in integral equation solvers, see Appendix~\ref{sec:MatrixDerivatives} for further details\footnote{Derivatives for tuning susceptances can be easily handled by knowledge of connected elements.}.

Losses can be incorporated by introducing loss matrix~$\RmatL$ into the equation~\eqref{eq:ImpedanceMatrix}, see~\cite{art:Capek2016OptCompModalCurrQ} for details.  In such a case, the total efficiency consists of two parts
\begin{equation}
\eta = \eta_\T{match}\eta_\T{rad},
\end{equation}
where $\eta_\T{match}$ describes the matching efficiency, and $\eta_\T{rad}$ is the radiation efficiency. The total efficiency value is strictly bounded by the maximal value of radiation efficiency~\cite{art:Capek2020FindOptTotalRefCoef}, and therefore, the maximal value of efficiency in the considered bandwidth is
\begin{equation}
\eta_\T{max} < 1.
\end{equation}
The final formula is then slightly more complex
\begin{equation}
\TARC\left(\omega\right) \approx \sqrt{1 - \eta_\T{max} -\frac{\omega_0^2}{2}\etaDD\bigg|_{\omega_0}\frac{\left(\omega - \omega_0\right)^2}{\omega_0^2}}. \label{eq:TARCApproxLossy}
\end{equation}
Nevertheless, the second derivative remains an indicator of bandwidth performance, and modified formulas~\eqref{eq:TotEffDDFull} and~\eqref{eq:FBWApproxQ} can be derived as well. Similarly, coupled generators can be used as well by using the maximum power transfer theorem as in~\cite{art:Broyde2022RadTransEfficiencies}.

\section{Example: Two Parallel Dipoles}
\label{sec:ExampleDipole2Port}
\begin{figure}[t]
\centering
\includegraphics[scale=1]{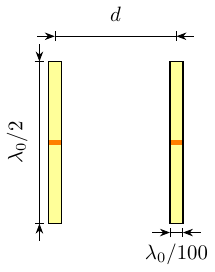}
\caption{Illustration of the considered antenna system. Orange lines represent the placement of feeding ports realized by discrete delta gaps~\cite{book:Gibson2021MoM}. All dimensions are related to the wavelength at $f_0$ in a vacuum. Parameter $d$ is used for various shifts between the two dipoles. Both dipoles are rotated by $90^\circ$ around their axis to allow lower $d$ values without overlapping.}\label{pic:Dipole2ParallelIllustration}
\end{figure}

The derived formulas are tested on two half-wave parallel dipoles made of a \ac{PEC} with varying separation distances. The voltages are chosen as in-phase~$\pVvec_\T{I}$ and out-of-phase~$\pVvec_\T{O}$, respectively, \ie{},
\begin{equation}
\begin{array}{cc}
\pVvec_\T{I} = \begin{bmatrix} 1 \\ 1 \end{bmatrix}, &
\pVvec_\T{O} = \begin{bmatrix} 1 \\ -1 \end{bmatrix}, 
\end{array}
\end{equation}
representing two canonical cases~\cite{book:BalanisAntTheory}. Line impedances~$\Lambmat$ and tuning elements $\pBmat_\T{L}$, either capacitors or inductors, are chosen to obtain $\eta\left(\omega_0\right) = 1$. The physical dimensions of the antenna system are related to wavelength~$\lambda_0$ at $\omega_0$ in vacuum, see Fig.~\ref{pic:Dipole2ParallelIllustration}.
In comparison,~\cite{art:Luomaniemi2021QfactMultiAntenna} proposed a multiport Q-factor that could be calculated directly from port quantities as
\begin{multline}
Q_\T{ZM} = \\ \omega_0\dfrac{\left\vert\pVvecH\dfrac{\partial\pGmatFS}{\partial \omega}\pVvec + \I\left(\pVvecH\dfrac{\partial\pBmatFS}{\partial \omega}\pVvec + \frac{1}{\omega_0}\left\vert\pVvecH\pBmatFS\pVvec\right\vert\right)\right\vert}{2\pVvecH\pGmatFS\pVvec}. \label{eq:QAalto}
\end{multline}
It can be shown that relation~\eqref{eq:QfactorTARCAdmittance} is equal to~\eqref{eq:QAalto} when the feeding vectors are the eigenvectors of the used matrices. As a definitive measure of the validity, the Q-factor evaluated from the fractional bandwidth~\eqref{eq:FBW} and formula~\eqref{eq:FBWApproxQ} is denoted as $Q_\T{FBW}$ and used whenever relevant as ground truth. 

\subsection{Increasing separation between dipoles}
\begin{figure}[t]
\centering
\includegraphics[width=0.99\columnwidth]{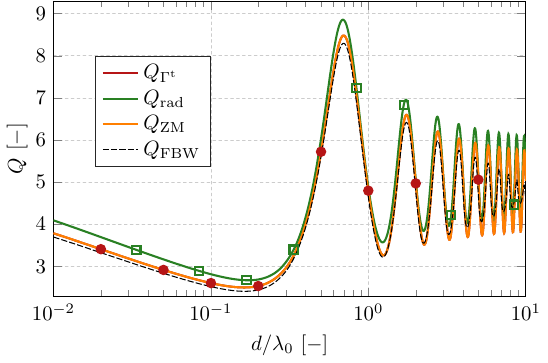}
\caption{Q-factor evaluated from different relations for in-phase feeding $\pVvec_\T{I}$. The spacing between dipoles is normalized to the wavelength. Value $Q_\T{FBW}$ represents the Q-factor computed from the fractional bandwidth~\eqref{eq:FBWApproxQ} at $\TARC_\T{max} = 0.2$ . Markers are used to highlight the curves.}\label{pic:Dipole2PortQfactorInPhase}
\end{figure}

\begin{figure}[t]
\centering
\includegraphics[width=0.99\columnwidth]{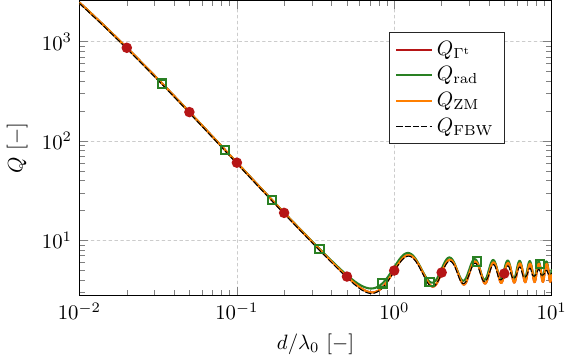}
\caption{Q-factor evaluated from different relations for out-of-phase feeding voltage $\pVvec_\T{O}$. The spacing between dipoles is normalized to the wavelength. Value $Q_\T{FBW}$  represents the Q-factor computed from the fractional bandwidth~\eqref{eq:FBWApproxQ} at $\TARC_\T{max} = 0.2$. Markers are used for highlighting the curves.}\label{pic:Dipole2PortQfactorOutOfPhase}
\end{figure}

Figures~\ref{pic:Dipole2PortQfactorInPhase} and~\ref{pic:Dipole2PortQfactorOutOfPhase} show a comparison of the mentioned formulas for the Q-factor. As discussed above, $Q_\TARC$ and $Q_\T{ZM}$ are identical. The curves representing $Q_\T{rad}$ show similar values as well. The Q-factor evaluated from the fractional bandwidth coincides with the results. Differences stem from the use of a derivative of the port admittance matrix instead of the \ac{MoM} impedance matrix. From a physical viewpoint, the resulting curves are consistent with it. Decreasing the distance between dipoles converges to a single dipole, for which out-of-phase feeding leads to conflicting currents canceling each other and leading to high Q-factor values. With an increase in distance, two dipoles begin to oscillate as the mutual coupling effect starts to slowly fade, agreeing with the conclusions made in~\cite{5778942}.

\subsection{Bandwidth approximation with Q-factor}
\begin{figure}[t]
\centering
\includegraphics[width=0.99\columnwidth]{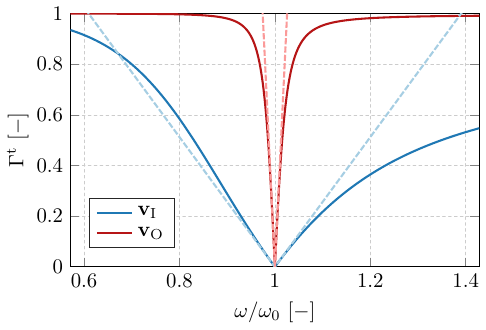}
\caption{Comparison of TARC~\eqref{eq:TARC} (solid dark line) and the first order approximation~\eqref{eq:TARCApprox} (dashed light line) with spacing between dipoles $d=\lambda_0/8$ for in-phase voltage vector~$\pVvec_\T{I}$ (blue curves) and out-of-phase voltage vector~$\pVvec_\T{O}$ (red curves).}\label{pic:Dipole2ParallelTARC}
\end{figure}

In Fig.~\ref{pic:Dipole2ParallelTARC}, the distance between dipoles is set to $d = \lambda_0/8$, and the \ac{TARC} bandwidth of in-phase and out-of-phase feeding is confronted with the true frequency dependence of \ac{TARC}. The frequency sweeps, with solid lines representing formula~\eqref{eq:TARC} and dashed lines representing the first-order approximation~\eqref{eq:TARCApprox}, clearly show the essential information obtained from the Q-factor about the fractional bandwidth of the connected feeding with respect to feeding voltages and corresponding input impedances. As mentioned in~\cite{art:Yaghjian2003ImpBandAnt}, the Q-factor prediction remains good for narrowband cases or low levels of $\TARC_\T{max}$.

\subsection{Bandwidth prediction from Q-factor}
\begin{figure}[t]
\centering
\includegraphics[width=0.99\columnwidth]{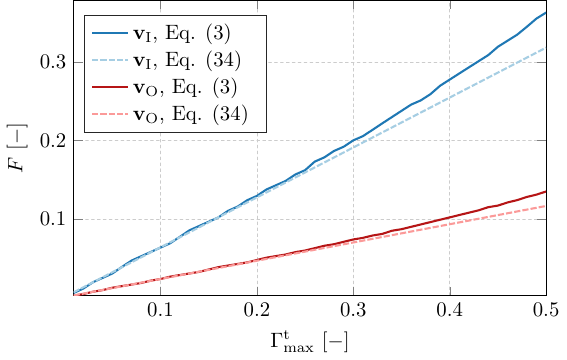}
\caption{Comparison of exactly computed fractional bandwidth~\eqref{eq:FBW} (solid lines) and prediction formula~\eqref{eq:FBWApproxQ} (dashed lines) for spacing between dipoles being $d = 3\lambda_0/4$.}\label{pic:Dipole2PortBandwidthApprox}
\end{figure}

Fractional bandwidth $F$ can be estimated from~\eqref{eq:FBWApproxQ}. Figure~\ref{pic:Dipole2PortBandwidthApprox} compares fractional bandwidth values calculated from~\eqref{eq:FBW} with estimation~\eqref{eq:FBWApproxQ} with respect to a given maximal value of \ac{TARC}. In this case, the spacing between both dipoles is set to $3\lambda_0/4$, and both sets of feeding vectors are used. It can be seen that formula~\eqref{eq:FBWApproxQ} is a good estimate up to high values of \ac{TARC}.

\section{Example: Five Parallel Dipoles}
\label{sec:ExampleDipole4Port}
\begin{figure}[t]
\centering
\includegraphics[scale=1]{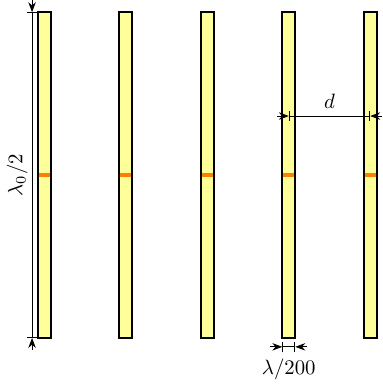}
\caption{Depiction of antenna array consisting of five parallel half-wave dipoles. Dimensions are included in the drawing. Orange lines represent the discrete delta gaps~\cite{book:Gibson2021MoM}.}\label{pic:Dipole5PortIllustration}
\end{figure}

Generally, Q-factors resulting from~\eqref{eq:QfactorTARCAdmittance} and \eqref{eq:QAalto} are not equal. This is demonstrated on an array of five parallel \ac{PEC} dipoles with equidistant spacing, non-uniform feeding coefficients, and complex matching networks. The size is related to free-space wavelength $\lambda_0$, see Fig.~\ref{pic:Dipole5PortIllustration}, and spacing $d$ is varied. Orange lines highlight the position of discrete feeding ports~\cite{book:Gibson2021MoM}. 
Three different feeding vectors~\cite{book:BalanisAntTheory} for broadside directivity are used for comparison:
\begin{enumerate}
\item triangle
\begin{equation}
\pVvec = \begin{bmatrix} 1 & 2 & 3 & 2 & 1 \end{bmatrix}^\mathrm{T},
\end{equation}
\item binomial
\begin{equation}
\pVvec = \begin{bmatrix} 1 & 4 & 6 & 4 & 1 \end{bmatrix}^\mathrm{T},
\end{equation}
\item and Dolph-Chebyshev\footnote{Considering $-20\,$dB sidelobes.}
\begin{equation}
\pVvec = \begin{bmatrix} 1 & 1.61 & 1.94 & 1.61 & 1 \end{bmatrix}^\mathrm{T}.
\end{equation}
\end{enumerate}

\begin{figure}[t]
\centering
\includegraphics[width=0.99\columnwidth]{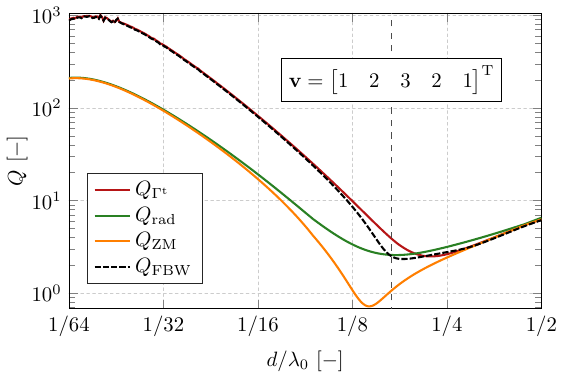}
\caption{The Q-factor evaluated from different relations for a triangle feeding vector with varying separations between dipoles. The black dashed curve represents the Q-factor computed from the fractional bandwidth~\eqref{eq:FBWApproxQ} at $\TARC_\T{max}=0.2$. The gray dashed line represents distance $d=\lambda_0/6$ where the \ac{TARC} is further analyzed. Red and black curves are slightly noisy as the result of imperfections in numerical modeling and matching networks optimization.}\label{pic:Dipole5PortQfactorTriangle}
\end{figure}

\begin{figure}[t]
\centering
\includegraphics[width=0.99\columnwidth]{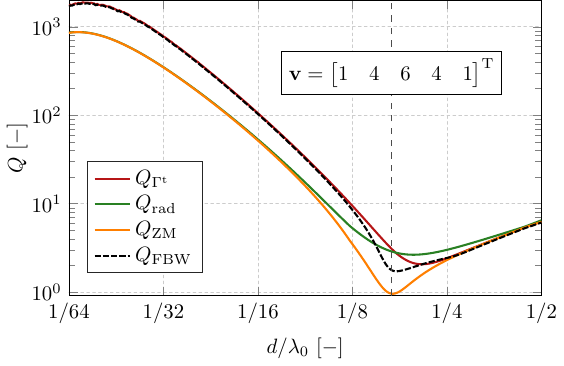}
\caption{The Q-factor evaluated from different relations for a binomial feeding vector with varying separations between dipoles. The black dashed curve represents the Q-factor computed from the fractional bandwidth~\eqref{eq:FBWApproxQ} at $\TARC_\T{max}=0.2$. The gray dashed line represents distance $d=\lambda_0/6$ where \ac{TARC} is further analyzed.}\label{pic:Dipole5PortQfactorBinomial}
\end{figure}

\begin{figure}[t]
\centering
\includegraphics[width=0.99\columnwidth]{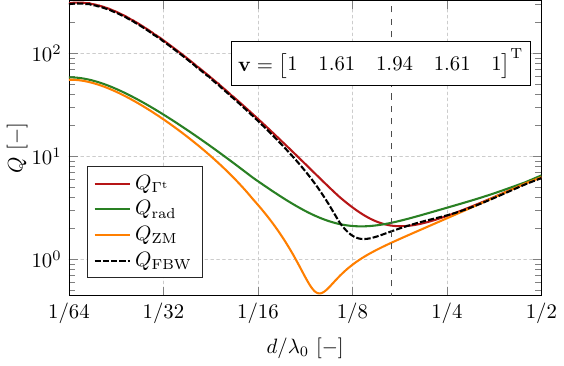}
\caption{Q-factor evaluated from different relations for the Dolph-Chebyshev feeding vector with varying separations between dipoles. The black dashed curve represents the Q-factor computed from the fractional bandwidth~\eqref{eq:FBWApproxQ} at $\TARC_\T{max}=0.2$. The gray dashed line represents distance $d=\lambda_0/6$ where \ac{TARC} is further analyzed.}\label{pic:Dipole5PortQfactorDolphChebyshev20}
\end{figure}

The above formulas~\eqref{eq:QfactorTARCAdmittance}, \eqref{eq:RadiationQPort}, \eqref{eq:QAalto} and~\eqref{eq:FBWApproxQ} are compared against $Q_\T{FBW}$ calculated at $\TARC_\T{max} = 0.2$ for three different feeding vectors in Figs.~\ref{pic:Dipole5PortQfactorTriangle},~\ref{pic:Dipole5PortQfactorBinomial} and~\ref{pic:Dipole5PortQfactorDolphChebyshev20}. All three cases become similar in Q-factor values for $d/\lambda_0> 1/4$, as the mutual coupling effect fades out. It is also evident that, at some point, the bandwidth predictions from \eqref{eq:RadiationQPort} and \eqref{eq:QAalto} become invalid, though they remain similar in value. Another issue is observable in the proximity of $d/\lambda_0 = 1/6$, where \eqref{eq:QfactorTARCAdmittance} and~\eqref{eq:FBWApproxQ} differ. 

\begin{figure}[t]
\centering
\includegraphics[width=0.99\columnwidth]{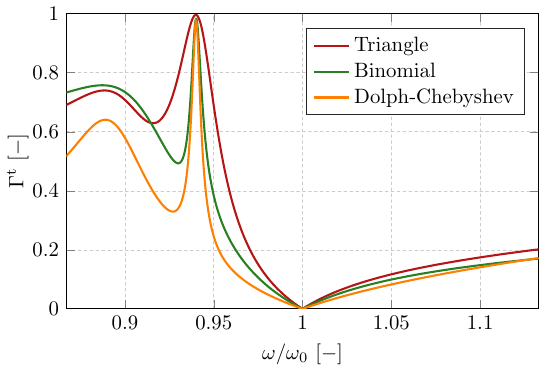}
\caption{Comparison of \ac{TARC} for three different feeding vectors. Distance between dipoles is $d = \lambda_0/6$.}\label{pic:Dipole5PortTARC}
\end{figure}

Further analysis indicates that the system is in a state of two closely spaced resonances. In such a case, Q-factor relations are no longer valid~\cite{art:Stuart2007LimitRelatQfactDoubRes}. Figure~\ref{pic:Dipole5PortTARC} shows the \ac{TARC} curves for the above-defined feeding vectors. In all three cases, the \ac{TARC} do not behave as a single separated resonance system, making the Q-factor prediction limited only for small values of $\TARC_\T{max}$.

\section{Example: Patch Arrays}
\label{sec:ExamplePatchArray}
The proposed theory is now applied to patch antenna arrays. This case can no longer be considered as electrically small. Nevertheless, patch antennas are well known to be narrowband, and the $Q$-factor provides a reliable estimate of the fractional bandwidth. This is demonstrated by two examples.

\subsection{Patch antenna array with four ports}
\begin{figure}[t]
\centering
\includegraphics[scale=0.74]{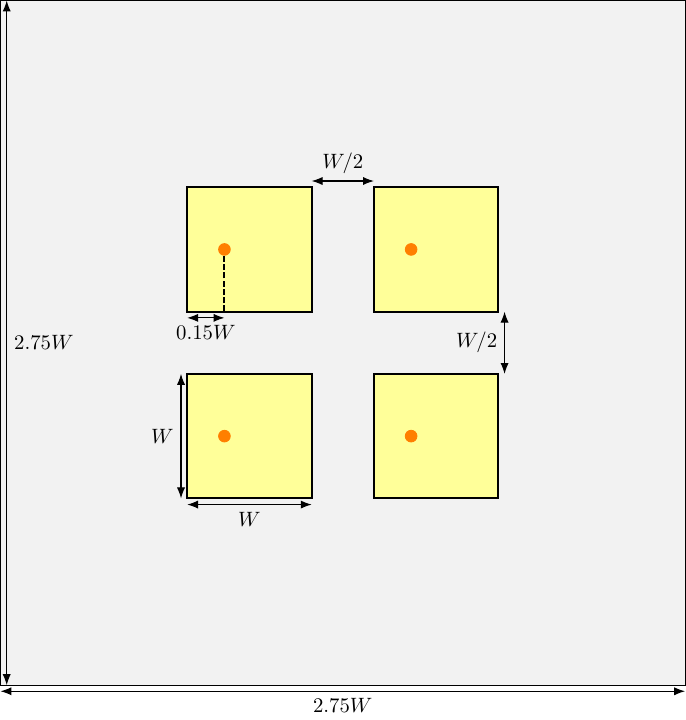}
\caption{Illustration of the considered antenna array with dimensions. Orange dots represent the port positions. The shaded area is the ground plane.}\label{pic:PatchArray4Port}
\end{figure}

The first array consists of four identical \ac{PEC} patch antennas placed in a vacuum $2\,$mm above the ground plane, as shown in Fig.~\ref{pic:PatchArray4Port}. All dimensions are related to wavelength $W \approx 0.94\lambda_0$ with the ground plane being of size $2.75W \times 2.75W$. 

\begin{figure}[t]
\centering
\includegraphics[width=0.99\columnwidth]{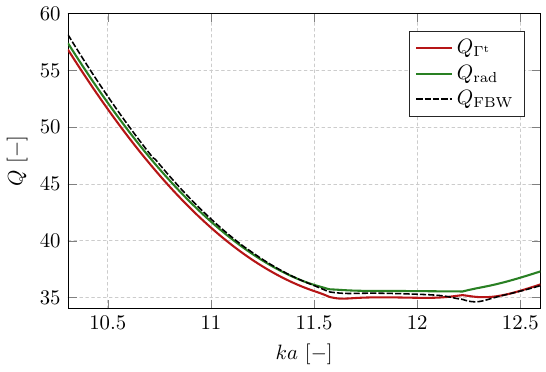}
\caption{Comparison of proposed formulas for Q-factor~\eqref{eq:RadiationQPort} and~\eqref{eq:QfactorTARCAdmittance} with formula~\eqref{eq:QAalto} derived in~\cite{art:Luomaniemi2021QfactMultiAntenna} for a four-element patch array. All ports are fed with uniform feeding and have the same matching at the examined frequency. The black dashed line represents the Q-factor computed directly from the fractional bandwidth.}\label{pic:PatchArray4PortQfactor}
\end{figure}

Figure~\ref{pic:PatchArray4PortQfactor} compares two proposed formulas for Q-factor with respect to electrical size $ka$. All ports were fed uniformly with identical line impedance and a tuning parallel element. Without a continuity treatment, the adjacent frequency samples could be matched with significantly different impedance values. Therefore, the line impedances and tuning elements were found to be continuous across frequency. In addition, the Q-factor computed from equation~\eqref{eq:FBWApproxQ} from the fractional bandwidth at level $\TARC_\T{max} = 0.2$ is included as well. Though the differences are more notable than in the first case of two dipoles, especially at the lower $ka$, all relations return similar results.  

\begin{figure}[t]
\centering
\includegraphics[width=0.99\columnwidth]{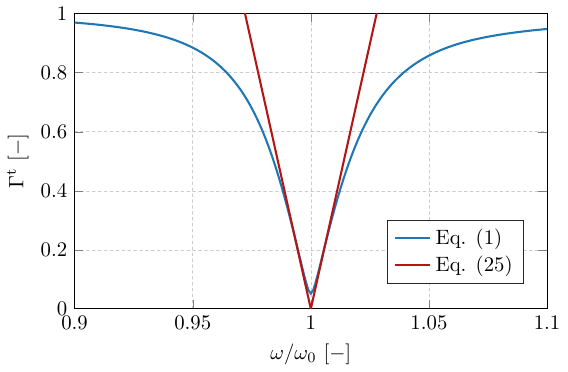}
\caption{Comparison of TARC~\eqref{eq:TARC} (blue line) and the first-order approximation~\eqref{eq:TARCApprox} (red line) for the four-port patch array.}\label{pic:PatchArray4PortTotEff}
\end{figure}

Considering antenna uniform feeding matched\footnote{In this case, the efficiency is equal to 0.997, causing a small deviation from the zero \ac{TARC}.} at $f_0$ corresponding to $\lambda_0$, the $\TARC$ and its approximation~\eqref{eq:QfactorTARCAdmittance} are compared in Fig.~\ref{pic:PatchArray4PortTotEff}. A good agreement is observed in the immediate vicinity of $f_0$.

\subsection{Patch antenna array with sixteen ports}
\begin{figure}[t]
\centering
\includegraphics[scale=0.95]{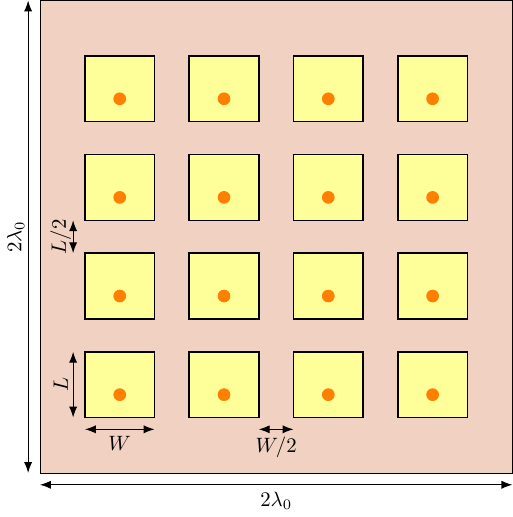}
\caption{Antenna array consisting of sixteen patches on substrate Astra MT77. Dimensions are $W = 36.77\,$mm and $L=34.77\,$mm. Each patch is fed with a discrete delta gap~\cite{book:Gibson2021MoM}, which is shifted by $5.27\,$mm from the center, and is highlighted by an orange dot. Ground plane and substrate dimensions are related to a free space wavelength $\lambda_0$ corresponding to $f_0$.}\label{pic:PatchArray16PortAntenna}
\end{figure}

As a last example, consider an array of sixteen copper patches designed on top of a $1.524\,$mm ultra low-loss material Astra MT77~\cite{dat:ASTRAMT77} with dielectric constant $\varepsilon_\T{r} = 3$ and dissipation factor $\tan\delta = 0.0017$. The design frequency is $f_0 = 2.4\,$GHz. The individual dimensions are listed in Fig.~\ref{pic:PatchArray16PortAntenna}.

\begin{figure}[t]
\centering
\includegraphics[width=0.99\columnwidth]{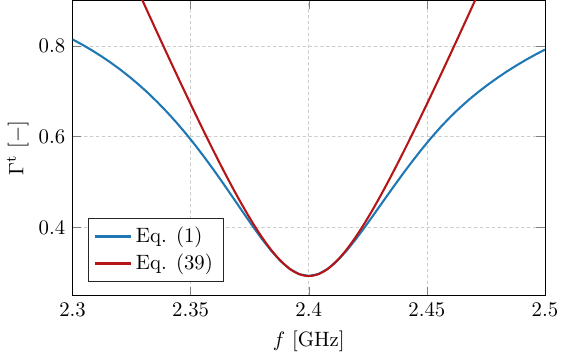}
\caption{Comparison of TARC~\eqref{eq:TARC} (blue line) and the first order  approximation~\eqref{eq:TARCApproxLossy} with incorporated losses (red line) for the patch array with sixteen ports.}\label{pic:PatchArray16PortTotEff}
\end{figure}

Using uniform feeding with optimal matching, the curves resulting from~\eqref{eq:TARC} and~\eqref{eq:TARCApproxLossy} are compared in Fig.~\ref{pic:PatchArray16PortTotEff}. The Q-factor again provides a reliable indicator of antenna bandwidth. 

%
\section{Conclusion}
\label{sec:Conclusion}
The theory of the multiport antenna Q-factor was introduced. Developed relations are directly related to the total active reflection coefficient and provide a fast way to estimate bandwidth from a single frequency sample, useful, for example, for structural optimization or feeding synthesis. The bandwidth dependence on the feeding coefficients was demonstrated through a series of examples, in which the proposed relations were compared with existing formulas, showing that they are special cases of the derived \ac{TARC} Q-factor. Generalized theory can be readily applied to a single-port case, yielding the well-known impedance Q-factor. Antenna array examples have shown that bandwidth estimation is not limited to electrically small antennas, provided the bandwidth is narrow and the resonance is isolated.

The open problems involve developing an effective optimization scheme that leverages knowledge of the multiport Q-factor, with variables including antenna shape, its feeding, and connected matching. The minor discrepancies found in the example consisting of four parallel dipoles, and the further analysis of the effect of matching, are out of the scope of this paper and are the subject of future research.

\appendices
\section{Port Matrix Derivatives}
\label{sec:MatrixDerivatives}
The existence of the analytical derivative of the impedance matrix~\cite{art:Vandenbosch2010ReactEnergyImp} can be utilized with the results of article~\cite{art:Capek2020FindOptTotalRefCoef} to find an analytical derivative of the port admittance matrix defined as
\begin{equation}
\pYmat_0 = \Pmat^\herm\Dmat^\herm\YmatFS\Dmat\Pmat.
\end{equation}
As only the $\YmatFS$ is frequency dependent, the derivative is
\begin{equation}
\frac{\partial \pYmatFS}{\partial \omega} = \Pmat^\herm\Dmat^\herm\frac{\partial\YmatFS}{\partial \omega}\Dmat\Pmat. \label{eq:YmatDerivative1}
\end{equation}
The derivative of the inverse impedance matrix reads~\cite{book:GolubMatrixComp}
\begin{equation}
\frac{\partial\YmatFS}{\partial \omega} = -\YmatFSMoM\frac{\partial \ZmatFSMoM}{\partial \omega}\YmatFSMoM,
\end{equation}
and substituting this result back to~\eqref{eq:YmatDerivative1} leads to
\begin{equation}
\frac{\partial\pYmatFS}{\partial \omega} = -\Pmat^\herm\Dmat^\herm\YmatFSMoM\frac{\partial \ZmatFSMoM}{\partial \omega}\YmatFSMoM\Dmat\Pmat.
\end{equation}
It is worth mentioning that the first $\YmatFS$ is not a Hermitian conjugate.

\bibliographystyle{IEEEtran}

\end{document}